\documentclass[twocolumn,showpacs,superscriptaddress,preprintnumbers,amsmath,amssymb,prb]{revtex4}

\usepackage{graphicx}
\usepackage{dcolumn}
\usepackage{bm}
\usepackage{ulem}
\usepackage{color}

\usepackage{epstopdf}

\begin{document}

\title{Discontinuous design of negative index metamaterials based on mode hybridization}

\author{Nian-Hai Shen}
\email[]{nhshen@ameslab.gov} \affiliation{Ames Laboratory and
Department of Physics and Astronomy, Iowa State University, Ames,
Iowa 50011, U.S.A.}

\author{Lei Zhang}
\affiliation{Ames Laboratory and Department of Physics and
Astronomy, Iowa State University, Ames, Iowa 50011, U.S.A.}

\author{Thomas Koschny}
\affiliation{Ames Laboratory and Department of Physics and
Astronomy, Iowa State University, Ames, Iowa 50011, U.S.A.}

\author{Babak Dastmalchi}
\affiliation{Ames Laboratory and Department of Physics and
Astronomy, Iowa State University, Ames, Iowa 50011, U.S.A.}
\affiliation{Center for Surface and Nanoanalytics, Johannes Kepler
University Linz, Altenbergerstrasse 69, 4040 Linz, Austria}

\author{Maria Kafesaki}
\affiliation{Institute of Electronic Structure and Laser, FORTH,
71110 Heraklion, Crete, Greece} \affiliation{Department of Materials
Science and Technology, University of Crete, 71003 Heraklion, Crete,
Greece}

\author{Costas M. Soukoulis}
\affiliation{Ames Laboratory and Department of Physics and
Astronomy, Iowa State University, Ames, Iowa 50011, U.S.A.}
\affiliation{Institute of Electronic Structure and Laser, FORTH,
71110 Heraklion, Crete, Greece}

\begin{abstract}
An electric inductor-capacitor (ELC) resonator provides a series of
electrical resonances and a pair of ELC resonators leads to the
split of each resonance into two modes, i.e., magnetic and electric
modes, corresponding to antisymmetric and symmetric current
distributions. With the meticulous design of the ELC resonator, we
can achieve a negative index metamaterial through mode hybridization
by overlapping the first electric resonance mode and the second
magnetic resonance mode. Such non-connected designs may offer
opportunities to achieve three-dimensional negative index
metamaterials.
\end{abstract}

\pacs{81.05.Xj, 73.20.Mf, 41.20.Jb, 78.20.Ci}

\maketitle

Metamaterials (MMs), formed by artificially designed, subwavelength
building blocks, have already brought us various, unexpected
phenomena and applications, such as negative refraction,
sub-diffractional imaging, and cloaking, etc. \cite{Shalaev2007,
Soukoulis2007, Zheludev2010, Soukoulis2011, Boltasseva,
Tassin2012NatPhoton} Up to now, in most of the
negative-index-metamaterial (NIM) designs, continuous metallic
components, are adopted to provide negative $\epsilon$ led by a
Drude-shape electric response, and negative $\mu$ is realized by
some specifically-designed magnetic resonator, which generally shows
a Lorentz-shape magnetic response. \cite{Shelby2001, Smith2000,
Huangfu2004, Chen2005, Zhou2006, Kafesaki2007, Guney2009} There also
have been some non-connected negative index metamaterials (NIMs)
achieved via engineering the electric and magnetic resonances of the
same plasmon mode to realize a spectral region, in which
permittivity and permeability are simultaneously negative. Zhou
\textit{et al.} \cite{Zhou2006OL} proposed to utilize the
interaction between the neighboring cut-wires in plane to lower the
electric resonance frequency so the band for negative permeability
can be settled within the region of negative permittivity since the
electric resonance is much stronger and much wider than the magnetic
one; they have both numerically and experimentally investigated a
possible design, ``H-shaped" cut-wire pair, which showed an exact
negative index band. \cite{Zhou2006APL} However, the disadvantage of
such a mechanism is the severe requirement of geometry parameters to
guarantee a strong enough interaction between neighboring unit cells
to shift the electric resonance frequency below the magnetic
frequency. Recently, Kante \textit{et al.} offered a scheme for
designing a non-connected negative index metamaterial by inverting
the hybridization of plasmon modes via misalignment of paired
cut-wires.\cite{Sellier2009OE, Kante2009PRB} However, it is easy to
see that the misalignment of cut wires will undoubtedly increase the
size of the unit cell and may hurt the ``subwavelength" property of
such metamaterials further.

In this paper, we propose a configuration to obtain a negative index
metamaterial with pure discontinuous elements, i.e., a pair of
electric-coupled inductor-capacitor (ELC)
resonator.\cite{Schurig2006apl, Padilla2007PRB} As well known, a
single ELC resonator provides a series of electric resonances, the
first two of which can be denoted as modes $|\omega_1\rangle$ and
$|\omega_2\rangle$ respectively. On the other hand, when two
original individual resonators compose a paired system, a
hybridization of plasmon modes \cite{Prodan2003Science,
Liu2009NatMater} for each individual resonator occurs, forming two
separate eigenmodes. Generally, the eigenmode at the lower frequency
is magnetic corresponding to antiparallel currents, while the
electric eigenmode with parallel currents is at a higher frequency
(i.e., $|\omega_{mi}\rangle < |\omega_{ei}\rangle$), where
\textit{i} represents the order of the resonance mode). Therefore,
an intuitive scheme for achieving a NIM is to adopt a paired system
of ELC resonators and try to overlap the spectral regions of
negative permittivity, $\epsilon$, of the first resonance and
negative permeability, $\mu$, of the second resonance.

\begin{figure}[htb]
 \centering
 \includegraphics[width=8cm]{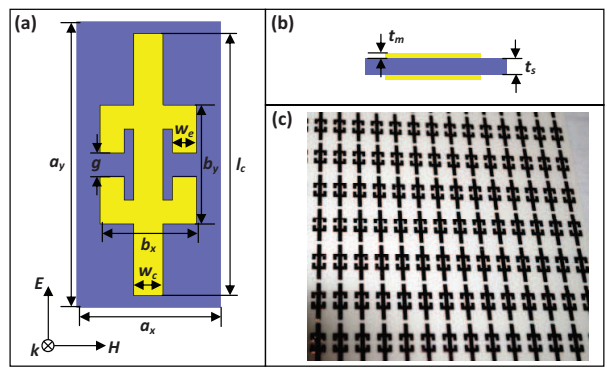}
 \caption{(Color online) (a) Front view and (b) top view to the schematic
of a unit cell of the discontinuous NIM design. (c) A photo of the
experimental sample. $a_x = 6$ mm, $a_y = 12$ mm, $b_x = 4$ mm, $b_y
= 5$ mm, $w_c = 1.2$ mm, $l_c = 11$ mm, $w_e = 1$ mm, $g = 1$ mm,
$t_s = 0.787$ mm, $t_m = 34$ $\mu$m.
 }
\end{figure}

Figure 1 presents a possible design of NIM, based on the
above-mentioned mechanism through a pair of ELC resonators. The ELC
resonator can be considered as a transformed E1 structure in Ref.
19, with the central bar elongated. Essentially, it is a combination
of an E1 structure and a cut wire. We will show below that such a
design makes negative $\epsilon$ for $|\omega_{e1}\rangle$ overlap
with negative $\mu$ for $|\omega_{m2}\rangle$ sufficiently for
negative \textit{n}. The schematic of mode-hybridization-mechanism
through a pair of ELC resonators for negative \textit{n} is
presented in Fig. 2, which also shows the surface current pattern of
each resonance mode specifically for our design. The different
thicknesses of the arrows schematically represent the different
strengths of the surface currents.

\begin{figure}[htb]
 \centering
 \includegraphics[width=8cm]{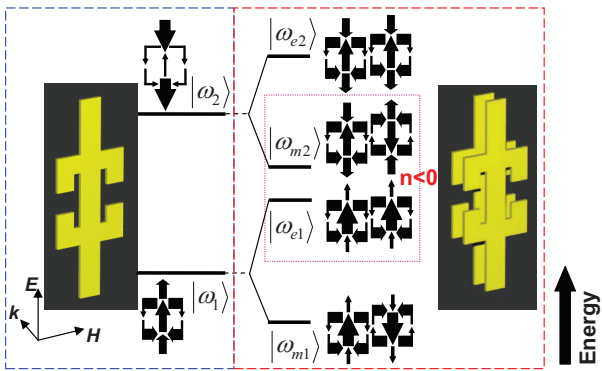}
 \caption{(Color online) Schematic of our proposed
 mode-hybridization mechanism to achieve negative index
 \textit{n}. Surface current patterns of our designed structure are
 shown next to the denotation of each mode.
 }
\end{figure}

We set our design to work in the gigahertz (GHz) region with the
corresponding parameters shown in the caption for Fig. 1. The
structure is patched on Rogers 4003 printed circuit board. A 0.787
mm-thick layer of dielectric ($\epsilon_s=3.55$ and loss tangent
0.0027) is coated with a 34 $\mu$m-thick copper layer on both sides.
We characterize the properties of the ELC resonator pair via both
numerical simulations and experimental measurements. The simulations
are performed with the commercial software package, CST Microwave
Studio. The software determines the reflection and transmission
information of the structure employing the finite integration
technique. In the simulations, only a single unit cell as shown in
Fig. 1(a) is taken into account with a periodic boundary condition
in \textit{x} and \textit{y} directions to mimic an array of ELC
resonator pair structure. Two ports are settled at each side of the
structure parallel to the \textit{x-y} plane, respectively. The
simulated reflection/transmission (R/T) spectra are shown as dashed
curves in Fig. 3(a) and (b). A photograph of the fabricated sample
for experimental study is shown in Fig. 1(c).

We measured the transmission and reflection properties of a
single-layer structure using an Agilent E8364B PNA network analyzer
supplied by a pair of Narda standard gain WR-62 horn antennas as
source and receiver. The working frequency is 12.4-18.0 GHz. Through
appropriate calibrations, we obtain clean spectra of reflection and
transmission as presented in Fig. 3 (a) and (b) with solid curves.
The overall qualitative agreement between simulations and
measurements is pretty good. To further demonstrate the properties
of this structure, we extract its electromagnetic parameters, i.e.,
\textit{n}, $\epsilon$, and $\mu$, from the obtained R/T information
by using the well-established retrieval procedure.
\cite{Smith2002PRB, Smith2005PRE, Koschny2005PRB} In the retrieval,
we assume the unit-cell size in \textit{z}-direction is 2.855 mm by
adding a 1 mm-thick air layer on each side of the structure.

\begin{figure}[htb]
 \centering
 \includegraphics[width=6cm]{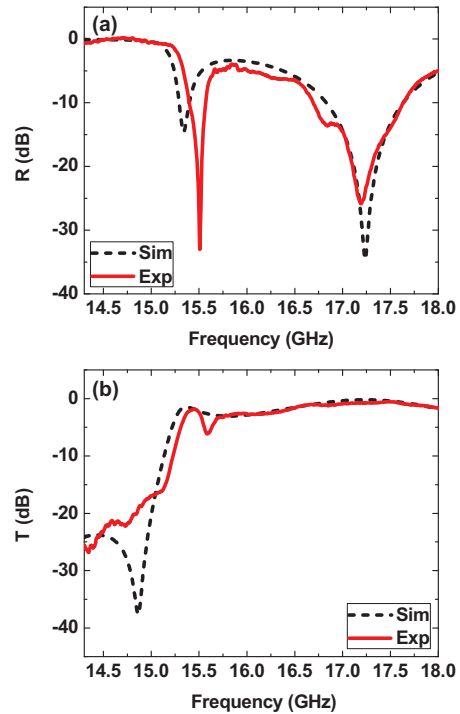}
 \caption{(Color online) Reflection (a) and transmission (b)
spectra for normal incident electromagnetic wave on the designed ELC
resonator pair structure: Simulations (black dashed curves) and
experiments (red solid curves).}
\end{figure}

\begin{figure}[htb]
 \centering
 \includegraphics[width=6cm]{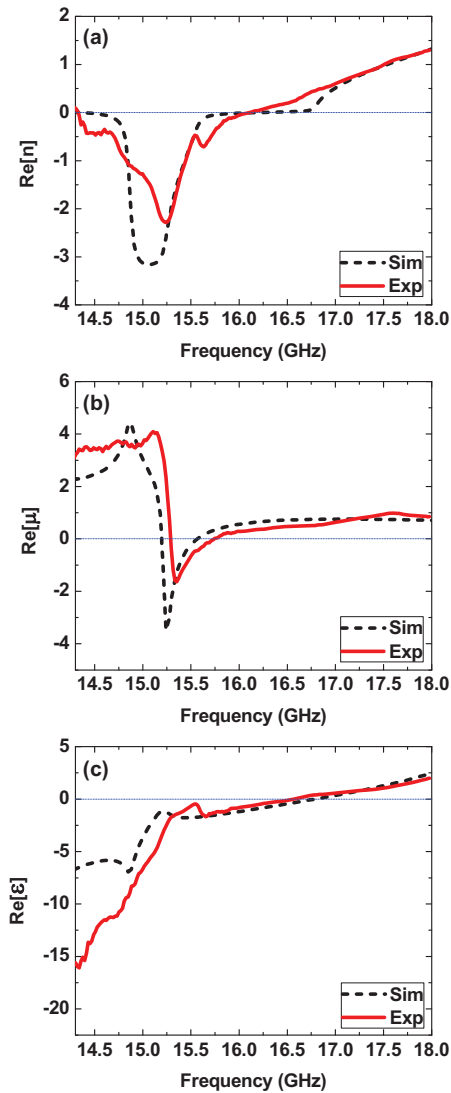}
 \caption{(Color online) Retrieved (a) refractive index
\textit{n}, (b) permeability $\mu$, and (c) permittivity $\epsilon$
of a single-layer of designed ELC resonator pair structure, using
the simulated (black dashed curve) and measured (red solid curve)
reflection and transmission data.}
\end{figure}

It is clearly seen from Fig. 4 the structure shows a narrow band of
negative \textit{n} with simultaneously negative $\epsilon$ and
$\mu$ around 15.5 GHz, both in simulation and measurement. The dip
value of refractive index \textit{n} by the experiment is as low as
$-2.4$, compared to $-3.1$ from the simulation. The simulated and
measured permeability $\mu$ dips to $-3.6$ and $-2$, respectively,
which originates from a strong magnetic resonance, the second order
magnetic resonance of our designed ELC resonator pair structure (the
first order magnetic resonance occurs at $\sim$6.6 GHz, not shown
here). The strength of the second magnetic resonance is comparable
to that of the first order, with quality factors of about 46 and 55,
respectively. Here, we would like to mention that, before any
optimization to the design, our structure has the maximum figure of
merit (FOM) about 10, which can be improved with further modulation
to the parameters. We also have completed a series of simulations by
changing $l_c$ in the structure (results not shown here). The
original separation of $\omega_{e1}$ and $\omega_{m2}$ is large with
a small $l_c$. When we increase $l_c$, $\omega_{e1}$ and
$\omega_{m2}$ become closer and finally, when $l_c=11$ mm, the
second magnetic resonance ($\mu<0$) overlaps with the first electric
resonance ($\epsilon<0$) giving negative \textit{n}. This confirms
our expectation of the mode hybridization mechanism for NIM designs.
In addition, the proposed design or various ELC resonator pair
structures, based on mode-hybridization mechanism, may render
negative index not only for GHz, but also in high frequency regions
and can be used to fabricate three-dimensional metamaterials with
negative index of refraction. \cite{Koschny2005PRB2}

In conclusion, pairing ELC resonators splits each resonance of the
original single structure into two modes, i.e., magnetic (lower
energy) and electric (higher energy) ones. We propose a mechanism to
realize negative index by hybridizing the first electric resonance
and the second magnetic resonance of a paired ELC resonator system.
Both numerical simulation and experimental measurements for an ELC
resonator pair design have confirmed our expectation for a band
showing a negative refractive index. Such a category of designs is
based on isolated components, which shall benefit the realization of
high dimensional negative index metamaterials. The idea may also
work in high frequency regimes, even at optical wavelengths with
various intriguing designs.

Work at Ames Laboratory was supported by the Department of Energy
(Basic Energy Sciences, Division of Materials Sciences and
Engineering) under contract No. DE-AC02-07CH11358. This was
partially supported by the European Community Project
NIM\rule[-2pt]{0.2cm}{0.5pt}NIL (Contract No. 228637).

\bibliographystyle{apsrev}

\newpage

\end{document}